\documentclass[prl,twocolumn]{revtex4-2}

\usepackage{graphicx}
\usepackage{amsmath}
\usepackage{bbold}
\usepackage{dcolumn}
\usepackage{bm}
\usepackage{array} 
\usepackage{xcolor}
\usepackage{colortbl} 
\usepackage{calc}
\usepackage{stackrel}
\usepackage{hyperref}
\newcommand{\halfquad}{\hspace{0.5em}}

\begin{document}

\preprint{APS/123-QED}

\title{Analytical Theory for Anomalous Diffusion in the Anderson Model with Heavy Tails}

\author{Elizaveta Safonova$^1$, Aleksey Lunkin$^2$ and Mikhail Feigel'man$^{2,3}$ }

\affiliation{$^1$Faculty of Mathematics and Physics, University of Ljubljana, Jadranska 19, 1000 Ljubljana, Slovenia, \\ $^2$Nanocenter CENN, Jamova 39, 1000 Ljubljana, Slovenia, \\ $^3$Jožef Stefan Institute, Jamova 39, 1000 Ljubljana, Slovenia}
\date{\today}

\begin{abstract}
We develop an analytical theory of anomalous transport in a noninteracting Anderson model with heavy-tailed hopping amplitudes. The broad distribution of hopping amplitudes gives rise to an
extended intermediate-time regime with a subdiffusive effective
exponent, despite the absence of interactions or genuine many-body
effects. By solving the transport equations analytically, we derive the time dependence of the mean-square displacement and identify a continuous crossover from an intermediate anomalous regime to asymptotically diffusive transport. As the localization transition is approached, the spatial extent of
the subdiffusive window diverges parametrically, while the crossover
to conventional diffusion remains finite in
units of $\Gamma_0^{-1}$.  This produces an increasingly broad anomalous
transport regime in space that can closely resemble Griffiths-type
transport observed near the many-body localization transition. Our results demonstrate that rare hopping processes alone provide a microscopic single-particle mechanism for robust transport anomalies, establishing an analytical benchmark for distinguishing interaction-induced effects from disorder-driven dynamics.
\end{abstract}

\maketitle

\textit{Introduction.--} The interplay between disorder and quantum coherence leads to localization phenomena in condensed matter systems. A paradigmatic example is Anderson localization \cite{andersonloc}, where strong disorder suppresses transport through destructive quantum interference, producing spatially localized single-particle states. Building on the earlier work of Fleishman and Anderson \cite{fleishmananderson}, who showed that short-range interactions do not necessarily destroy localization, the many-body localized (MBL) phase \cite{BASKO20061126} provides an example of an interacting insulating state characterized by the absence of transport \cite{NandkishoreHuse2015}. Coupling to external degrees of freedom, such as phonons, can nevertheless restore transport through mechanisms including variable-range hopping \cite{Mott01041969}.

Considerable attention has focused on the thermal side of the MBL transition, where numerical studies have reported anomalously slow relaxation and subdiffusive transport \cite{barlev2014,barlev2015,Sierant2025,spintransport, SubdifAnderson}. A widely accepted explanation invokes Griffiths effects \cite{Agarwal2015} arising from rare insulating regions embedded in an otherwise conducting phase. Their exponentially large resistances, combined with exponentially small occurrence probabilities, generate broad distributions of relaxation times and power-law dynamical correlations \cite{Vosk2015}, providing a natural mechanism for subdiffusion.

Whether subdiffusion persists asymptotically or merely reflects a long crossover remains an open question \cite{Luitz2016,Sierant2025}. Numerical methods, including exact diagonalization and tensor-network simulations, are restricted to finite system sizes and evolution times, making it difficult to distinguish genuine long-time dynamics from finite-size effects \cite{barlev2015,Sierant2025}. While many works report power-law transport with continuously varying dynamical exponents \cite{barlev2015,Luitz2016,Agarwal2015, SubdifAnderson}, others argue that the observed subdiffusion is a long-lived transient that ultimately crosses over to ordinary diffusion \cite{Sierant2025}. The existence of a true asymptotic subdiffusive phase therefore remains an open question.

These limitations motivate the search for effective models that capture anomalous transport without the finite-size restrictions of interacting many-body systems. Random matrix theory provides a natural framework for describing disordered systems and, in suitable regimes, serves as an effective description of Anderson-type models with extended or critical states \cite{FyodorovMirlinSommers,MIRLIN2000}. In particular, random hopping models with algebraically distributed amplitudes provide a minimal setting for studying slow dynamics induced by long-range couplings while isolating disorder effects from genuine many-body correlations~\cite{BOUCHAUD1990127,RosenzweigPorterOriginal}.

Hamiltonians with power-law-distributed off-diagonal matrix elements have previously been studied in zero-dimensional random matrix ensembles \cite{EversMirlin2008, TarquiniTarziaLevy, PhysRevB.106.094204}. Some of these models exhibit nontrivial spectral statistics and multifractal eigenstates \cite{RME, Khaymovich2015Multifractality, Khaymovich2020FragileExtended, Khaymovich2021DynamicalPhases}. However, the absence of spatial structure limits their applicability to transport problems. Using the framework developed in Ref.~\cite{FyodorovMirlinSommers}, we map these ideas onto a real-space representation using the Lévy–Rosenzweig–Porter ensemble \cite{SizeauBouchaud, TarziaBiroliLRP}, which has been extensively studied in zero-dimensional settings in recent years \cite{SpectralLRP,DensCorrLevy,LunkinTikhonov}. This construction allows us to investigate whether anomalous transport and power-law dynamics, often associated with Griffiths physics in interacting systems, can already emerge in an effective noninteracting setting.

\begin{figure}[!h]
    \centering
    \includegraphics[width=\linewidth]{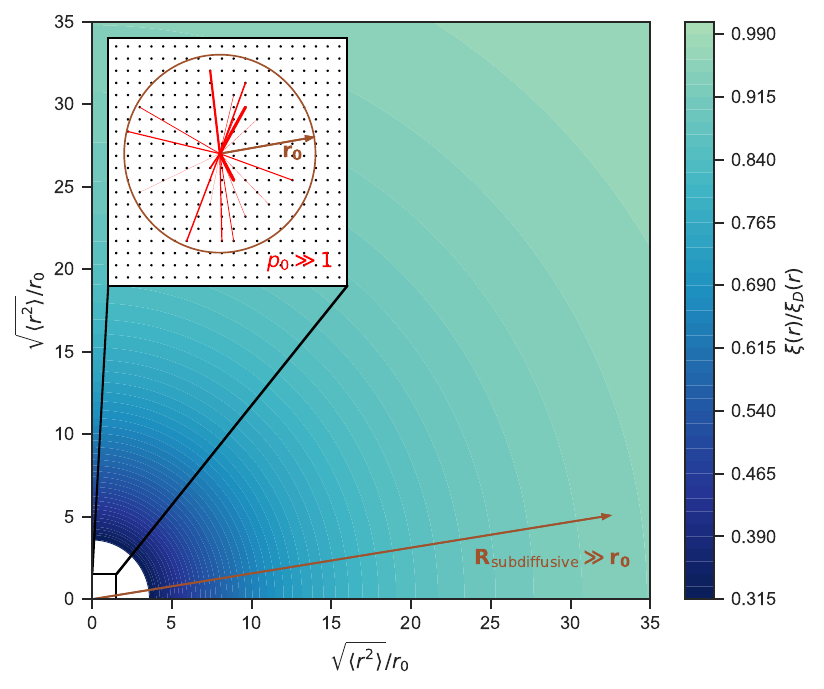}
    \caption{Color map of the normalized scaling exponent $\xi(r)$ as a function of 
propagation distance for $\mu=1.03$, where 
$\xi_D(r) = 1$. The figure demonstrates that 
the subdiffusive regime extends over distances much larger than the mean free path $r_0$. The white region indicates the absence of data. The inset shows a microscopic illustration of the model, where each site is, on average, 
connected to $p_0 \gg 1$ neighboring sites within a distance $r_0$, forming a network with multiple broadly distributed connections to nearby sites. The thickness of each red line represents the amplitude of the corresponding hopping term $t_{nm}$.} \label{Fig1}
\end{figure}

In this Letter, we develop a theory of subdiffusive transport in the delocalized phase near the Anderson localization transition. We show that the mean-square displacement $\langle r^2(t)\rangle$ grows \textit{sublinearly} in $t$ over a broad range of intermediate times. Such subdiffusive
behavior extends over a spatial scale $R_{\rm sub}$ that greatly exceeds the microscopic hopping length $r_0$ (Fig.~\ref{Fig1}).

\begin{figure}[!t]
    \centering
    \includegraphics[width=\columnwidth]{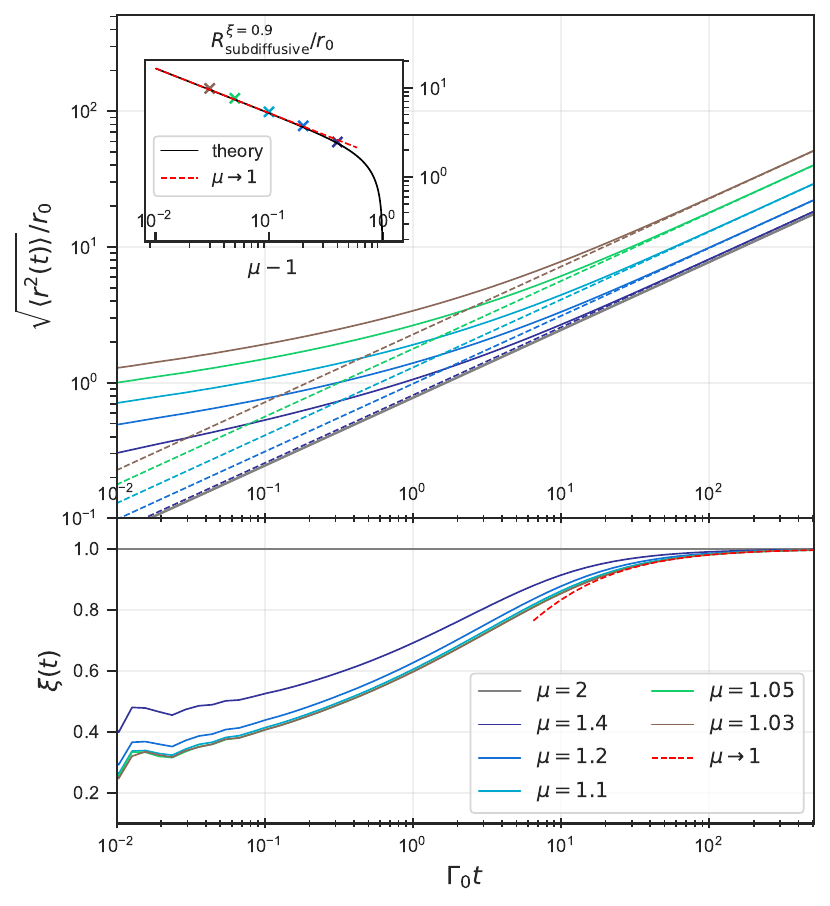}
    \caption{
Top: Typical displacement $r_{\text{typ}}(t)$, normalized over hopping range $r_0$, obtained from Eqs.~(\ref{K func final}, \ref{lambda c defs}, and \ref{r^2}) for various values of $\mu$. Dashed lines show the long-time diffusive asymptotic $r_{\text{typ}}\propto \sqrt{t}$. Bottom:  effective exponent $\xi(t)$, approaching the diffusive value $\xi=1$(gray) at long times, the red dashed line is the $\mu\to1$ asymptotic result, \eqref{xi_limit}. Inset to the top panel: power-law increase of $R_{\text{sub}}^{\xi=0.9}$ upon approaching the localization transition $\mu\rightarrow 1$. Crosses: numerical evaluation of the full theory; lines:
asymptotic prediction \eqref{R subdiff limit}.}
\label{Fig2}
\end{figure}

As the localization transition is approached, $R_{\rm sub}$ diverges parametrically as $R_{\rm sub} \sim r_0/\sqrt{\mu-1}$, where the control parameter $\mu$ approaches its critical value $\mu=1$ (inset of Fig.~\ref{Fig2}), resulting in a long-lived crossover regime. At asymptotically long times the system ultimately recovers conventional diffusion, so that the observed 
subdiffusive behavior does not represent a new asymptotic phase but a parametrically extended crossover. During this regime, transport is characterized by a continuously evolving effective exponent,
\begin{equation}\label{xi def}
    \xi \equiv \frac{d\log{\langle r^{2}(t)\rangle}}{  d\log t},
\end{equation}
which approaches the diffusive value $\xi=1$ at long times (Fig.~\ref{Fig2}). We emphasize that ``subdiffusive'' here refers to the local
logarithmic exponent $\xi(t)<1$. The mean-square displacement itself
can remain above the extrapolated long-time diffusive law. To avoid localization corrections, we focus primarily on lattices with $d>2$. These results are quantitatively consistent for both the orthogonal ($\beta=1$) and unitary ($\beta=2$) symmetry classes. The analytical predictions are further tested by numerical evaluation of the full theory and by direct simulations of the microscopic Hamiltonian \eqref{anderson hamiltonian}.

\textit{Model and conductivity correlation function.--}
To investigate transport near the localization transition, we consider a class of disordered tight-binding models with finite-range hopping and broadly distributed hopping amplitudes. By continuously interpolating between Gaussian and Lévy statistics of the hopping amplitudes, this framework enables us to study transport from the conventional diffusive regime to the vicinity of Anderson localization through the statistics of the off-diagonal disorder.

We consider the Anderson Hamiltonian on a $d$-dimensional cubic lattice,
\begin{equation}\label{anderson hamiltonian}
{\cal H}
=
\sum_n u_n a_n^\dagger a_n
+
\sum_{n,m} t_{nm} a_n^\dagger a_m .
\end{equation}
The on-site energies $u_n$ are independent random variables drawn from a broad distribution $P_D(u)$ with characteristic width $W$, the largest energy scale in the problem. The hopping matrix elements are distributed according to
\begin{equation}\label{hopping distr}
{\cal P}(t_{nm})
=
(1-p_{nm})\delta(t_{nm})
+
p_{nm}P(t_{nm}),
\end{equation}
where $p_{nm}$ is determined by the real-space profile as
$p_{nm}
=
\tilde p\,w(\mathbf r_n-\mathbf r_m)$, where $w(\mathbf r)=\theta(r_0-|\mathbf r|)$ restricts hopping to a finite range $r_0$. Defining $w_0\equiv\sum_{\mathbf r}w(\mathbf r)$ as the number of sites within this range, the connection probability is
$
\tilde p=\frac{p_0}{w_0},
$
where $p_0\gg1$ is the mean coordination number.
 Thus, sites separated by more than $r_0$ are not
directly coupled, whereas sites within this range are connected with
probability $\tilde p$, in which case the hopping amplitude is drawn
from the distribution $P(t_{nm})$. This model was introduced in
Ref.~\cite{FyodorovMirlinSommers}. For simplicity, we choose the sharp
cutoff $w(\mathbf r)=\theta(r_0-|\mathbf r|)$, which provides a
transparent physical picture. However, none of our results
depend on this specific choice of $w(\mathbf r)$: the same physics
arises with any bounded function that has a finite second moment.

Throughout this work, we choose the hopping amplitudes to obey the following power-law distributions:

\begin{center}
\begin{tabular}{|c|c|}
\hline
$\beta=1$ & $\beta=2$ \\
\hline
$t_{nm}=t_{mn}$ &
$t_{nm}=t_{mn}^*=|t_{nm}|e^{i\theta_{nm}},\quad
\theta_{nm} \sim\mathcal U(0,2\pi)$
\\
\hline
\multicolumn{2}{|c|}
{$
P_{\mu}(|t|)d|t|
\simeq
\frac{h_0^\mu\mu\,d|t|}
{p_0\Gamma(1-\mu/2)|t|^{1+\mu}}
$}
\\
\hline 
\end{tabular}
\end{center}
where $\mathcal U(0,2\pi)$ denotes the uniform distribution and $h_0$ sets the characteristic energy scale of the hopping elements.

The parameter $\mu$ controls the statistics of rare large hopping amplitudes and therefore constitutes the key parameter of the model. It provides a continuous interpolation between transport regimes. At $\mu=2$ the system exhibits conventional diffusion with a finite diffusion coefficient, while decreasing $\mu$ enhances quantum interference and drives the system toward the Anderson localization transition at $\mu=1$.

The statistical properties of the ensemble are insensitive to the microscopic form of $P(t_{nm})$ and
depend only on its asymptotic power-law tail. This universality follows
from the convergence of broad classes of heavy-tailed distributions to
the Lévy stable distribution $L_\alpha (x)$,
\begin{equation}\label{Levy def}
\int_0^\infty L_{\mu/2}(x)e^{-\lambda x}\,dx
=
e^{-\lambda^{\mu/2}},
\end{equation}
whose form is determined solely by the tail exponent $\mu$. The Gaussian limit is recovered continuously at $\mu=2$, where $L_1(x)=\delta(x-1)$.

To characterize transport we study the disorder-averaged retarded-advanced correlation function
\begin{equation}
K_{jk}(\kappa)=
\overline{
\left\langle j\left|
\frac{1}{E+\frac{i\kappa}{2}-\hat H}
\right|k\right\rangle
\left\langle k\left|
\frac{1}{E-\frac{i\kappa}{2}-\hat H}
\right|j\right\rangle},
\label{K func}
\end{equation}
where $\kappa$ is an imaginary-frequency variable, $i\kappa\equiv\omega+i0$, and the overline denotes averaging over the disorder in $\hat H$. In this formulation, the diagonal elements of $\hat H$ correspond to the
on-site random energies $u_i$, while the off-diagonal elements are drawn
from the distribution $P(t_{ij})$. Thus, the spatially dependent problem
can be mapped onto the so-called Lévy--Rosenzweig--Porter (LRP) matrix
ensemble~\cite{TarziaBiroliLRP,SpectralLRP,DensCorrLevy}, where the
diagonal entries of $\hat H$ are random variables broadly distributed over
a characteristic width $W$, the largest energy scale in the problem. The
mean level spacing, $ \Delta=(w_0 P_D(E))^{-1}\sim W/(r_0/a)^d,$
is therefore the smallest energy scale ($a$ is the lattice spacing). 
The correlation function~\eqref{K func} is directly related to the frequency-dependent conductivity~\cite{WEGNER198015}. At zero temperature and low frequencies ($\omega\ll W$), only the retarded-advanced contribution survives,
\begin{equation}
\label{conductivity main}
\sigma(\kappa)
=
-\frac{e^{2}\kappa^2}{4dv\pi}
\sum_k
(\mathbf r_j-\mathbf r_k)^2
K_{jk}(\kappa)
=
\frac{e^{2}\kappa^2}{2v\pi}
\frac{d^2\hat{K}_q(\kappa)}{d q^2}\Big|_{q=0},
\end{equation}
with $\hat{K}(q)$ assumed to be spherically symmetric. For sufficiently small frequencies and momenta, the correlation function must recover the conventional diffusive form,
\begin{equation}
\hat K_q(\kappa)
=
\frac{2\pi P_D(E)}
{D q^2+ \kappa},
\label{K(q) diffusive form}
\end{equation}
which serves as the reference point for the anomalous transport theory developed below.

\textit{Physical discussion and regime of validity.--}
The present model provides a framework for extending the supersymmetric
approach developed for power-law random matrices to spatially extended
systems. Previous studies of these ensembles were primarily restricted to
zero-dimensional sigma models. In addition to the two standard energy scales, the spectral width $W$ and the level spacing $\Delta$, Rosenzweig--Porter models naturally contain a third energy scale, $\Gamma_0$ (defined in Eq.~\eqref{Gamma}) \cite{SkvKrv,TarziaBiroliLRP}. For the particular case of the LRP model, it has an exact solution given in \cite{DensCorrLevy, LunkinTikhonov}. However, in the spatially dependent problem, it includes the parameter $p_0$ 
that measures the average number of available sites within the hopping range $r_0$. 
A characteristic feature of the Lévy--Rosenzweig--Porter ensemble is the statistical
independence of the real and imaginary parts of the self-energy $\Sigma(E)$. Physically,
this reflects the fact that $\Re\Sigma(E)$ receives contributions from a wide energy range
of order $W$, while $\Im\Sigma(E)$ is governed only by the local environment of the level
at energy $E$. As shown in Ref.~\cite{SkvKrv}, this separation of scales leads to
essentially independent fluctuations of the two quantities. The distribution of $\Re\Sigma$ was obtained in Ref.~\cite{SpectralLRP}. Since it merely induces a small renormalization of the spectrum, of relative order $h_0/W$, its effect can be neglected in the present problem. By contrast, the imaginary part is distributed according to a Lévy distribution, defined by a characteristic energy scale $\Gamma_0$.

A more physical interpretation of $\Gamma_0$ is that it sets the characteristic energy scale of wave-function hybridization. Levels separated by less than $\Gamma_0$ are strongly mixed and exhibit Wigner--Dyson level repulsion, whereas levels separated by more than $\Gamma_0$ are effectively independent, owing to the suppression of the overlap between their eigenfunctions. Thus, $\Gamma_0$ defines the energy window within which eigenstates are significantly hybridized. This leads to the condition $\Delta \ll \Gamma_0 \ll W$.

The mean-field approximation requires a large number of sites within the
hopping range. Defining $w_0=\sum_{\mathbf r}w(\mathbf r)$ as
the number of sites inside the hopping volume, we would have $w_0 \sim (r_0/a)^d\gg1$
as the mean coordination number for a fully-connected model with $\tilde p = 1$. The parameter $r_0$ therefore plays the role of an effective microscopic length scale,
similar to the mean free path in a conventional diffusion problem. 

In addition, the dimensionless conductance must be parametrically large
to suppress localization corrections. This requirement takes the form $P_D(E)\Gamma_0\,w_0\gg1 .$ Together with the weak hybridization condition $P_D(E)\Gamma_0\ll1$, it defines an intermediate regime governed by the competition between weak level hybridization and large connectivity required for the mean-field approximation. For the existence of this regime, a strong inequality $w_0 \gg 1$ must hold.

The localization limit $\mu\to 1$, however, is subject to additional constraints.
As discussed above, the theory requires $\Gamma_0\ll W$. At the same time,
the definition of $\Gamma_0$ in Eq.~\eqref{Gamma} contains a $\mu$-dependent
coefficient $a(\mu)$ that diverges as $\mu\to 1$, thereby invalidating this
condition sufficiently close to the localization limit. The range of validity
can be extended by tuning $P_D(E)h_0\to 0$; nevertheless, there always remains
a threshold $\mu_*>1$ such that the theory is valid only for $\mu>\mu_*$.

\textit{Subdiffusion as a long intermediate regime.--}\label{Results} We present results for the mean-square displacement $\langle r^2(t)\rangle$, demonstrating a long intermediate regime before diffusion sets in. We use the supersymmetric approach originally described in Ref.~\cite{FyodorovMirlinSommers}, together with previous studies of the zero-dimensional LRP model~\cite{DensCorrLevy,LunkinTikhonov} (see End Matter for the derivation). Figures~1--3 present numerical evaluations of the analytical theory,
while Fig.~4 provides an independent check by direct time-evolution
simulations of the Anderson Hamiltonian. The conductivity $\sigma(\omega)$ is related to the mean-square displacement $\langle r(t)^2 \rangle$ by Eq.~\eqref{conductivity main}. The main result of our calculations is the
correlation function $\hat{K}(q,\kappa)$ for arbitrary imaginary frequencies $i\kappa$: 
\begin{equation}\label{K func final}
    \hat{K}_q (\kappa) = \frac{c_0(\kappa)}{1-\tilde{w}(q) \lambda_0(\kappa)},
\end{equation}
where $\tilde{w}(q) = \sum_{\bm{r}} w(\bm{r})e^{i \bm{qr}}$.  We introduce the notation $V_\alpha(\kappa) \equiv \left[\frac{\Gamma_{0}}{\Gamma_{\kappa}}\right]^{\alpha}\Gamma(\alpha)\int dr\frac{L_{\mu/2}\left(r\right)}{\left(\frac{\kappa}{\Gamma_{\kappa}}+r\right)^{\alpha}}$, where $L_{\mu/2}(r)$ is the Lévy distribution function defined in Eq.~\eqref{Levy def};
then the functions $c_0(\kappa)$ and $\lambda_0(\kappa)$ can be represented as follows:
\begin{equation}\label{lambda c defs}
    c_0(\kappa) = \frac{2 \pi P_D(E)\left[V_{\mu/2}(\kappa)\right]^{2}}{\Gamma_{0} V_{\mu-1}(\kappa)},\quad    \lambda_0(\kappa) =  \frac{1}{w_0}\frac{\frac{\mu}{2} V_{\mu-1}(\kappa)}{\Gamma\left(2-\frac{2}{\mu}\right)} 
\end{equation}
where $\Gamma_\kappa$ is given by the solution of the self-consistent equation

\begin{equation}\label{Gamma}
\left(\frac{\Gamma_{\kappa}}{\Gamma_{0}}\right)^{\mu/2}=\frac{\frac{\mu}{2} V_{\frac{\mu}{2}-1}(\kappa)}{\Gamma\left(1-\frac{2}{\mu}\right)},\quad  \left[\frac{\Gamma_{0}}{ h_0}\right]^{\mu-1} =  P_D(E)h_{0} a(\mu).
\end{equation}
Here $a(\mu)=\frac{\sqrt{\pi}\Gamma\left(\frac{\mu-1}{2}\right) \Gamma\left( 2 - \frac{2}{\mu}\right)}{[\Gamma\left(\mu/2\right)]^2\cdot 2^{1-\mu}}$. The parameter $\Gamma_0$ is the $\mu$-dependent energy scale discussed above. We use it as the natural energy scale and $\Gamma_0^{-1}$ as the corresponding time scale. Note that in the zero-frequency limit $\kappa \rightarrow i0$, $\Gamma_\kappa$ equals $\Gamma_0$.

To illustrate our results for anomalous diffusion, we calculate the time-dependent mean-square displacement $\langle r^2(t)\rangle$.  
\begin{equation}\label{r^2}
\langle r^2(t) \rangle =   \frac{d}{2 \pi P_D(E)}\int \frac{d\omega}{2\pi}(1-e^{-i\omega t})\frac{d^2}{d q^2}\hat{K}\left(q\right)\biggr|_{q=0} .
\end{equation}

It is useful to introduce the notation for the typical length $r_{\mathrm{typ}}(t) = \sqrt{\langle r^2(t) \rangle }$.
For $\mu=2$, the correlation function \eqref{K func final} has a standard diffusive form \eqref{K(q) diffusive form}, giving 
$\langle r^2 (t)\rangle = 2dDt$ with diffusion coefficient $D = \Gamma_0 \frac{w_2}{w_0}$, where $w_2 = \frac{1}{2d}\sum\limits_{\bm{r}}w(|\bm{r}|) r^2$ and $w_0 = \sum\limits_{\bm{r}} w(|\bm{r}|)$.
For any other $\mu<2$, the diffusive limit in Eq.~\eqref{K(q) diffusive form} is recovered after a crossover time that is significantly longer than $\Gamma_0^{-1}$ (see the top panel of Fig.~\ref{Fig2}). In this case, the diffusion coefficient is $\mu$-dependent.
\begin{equation}\label{D coef}
    D = \frac{\mu}{2} \Gamma\left(2 - \frac{2}{\mu}\right)\Gamma_0 \frac{w_2}{w_0}.
\end{equation}
The typical displacement after time $t$ grows as $r_{\textrm{typ}}(t)
\sim \left(\Gamma_0 t\right)^{\xi(t)/2}$, where $\xi(t)$ is defined in \eqref{xi def}. 
In the long-time limit $t\Gamma_0 \gg 1$, an approximate asymptotic form is derived:
\begin{equation}\label{correction <r2>}
    \langle r^2 (t)\rangle = 2dDt + d\frac{w_2}{w_0}\frac{\mu(2-\mu)}{\mu-1}.
\end{equation}
This allows us to express the exponent $\xi(t)$ in a simple asymptotic form:
\begin{equation}\label{xi_limit}
    \xi(t) = \frac{\Gamma_0 t}{ \Gamma_0 t + \frac{(2-\mu)}{\Gamma\left(2-2/\mu\right) (\mu-1)}} \stackrel{\mu\rightarrow 1}{=}\frac{\Gamma_0 t}{\Gamma_0 t + 2} .
\end{equation}

For completeness, Eq.~(14) also allows us to characterize the
crossover time explicitly. Defining $t_\xi$ by
$\xi(t_\xi)=\xi$, with a fixed $\xi<1$, we obtain
\begin{equation}
    \Gamma_0 t_\xi =
    \frac{2-\mu}
    {\Gamma\!\left(2-\frac{2}{\mu}\right)(\mu-1)}
    \frac{\xi}{1-\xi} \xrightarrow[\mu\to1^+]{}
    \frac{2\xi}{1-\xi}.
\end{equation}
The asymptotic approach $\xi(t) \to 1$ as $t \to \infty$ allows us to define a length scale
$R_{\text{sub}}^\xi = r_{\text{typ}}(t_\xi)$.
This length scale characterizes a  crossover length between the sub-diffusive and standard diffusion regimes:
\begin{equation}
\label{R subdiff limit}
    R_{\text{sub}}^\xi =\sqrt{\frac{d w_2}{w_0} \frac{\mu(2-\mu)}{\mu-1} \frac{\xi}{1-\xi}} \stackrel{\mu\rightarrow 1}{\sim} \frac{r_0}{\sqrt{\mu-1}}.
\end{equation}

It is important that for any fixed $\xi<1$ the crossover scale
$R^\xi_{\rm sub}$ becomes much larger than $r_0$ as
$\mu\to1^+$, as illustrated in Fig.~2. In contrast, the
corresponding crossover time $t_\xi$ remains finite when measured
in units of $\Gamma_0^{-1}$. The singular behavior near the
localization transition therefore manifests itself as a
parametrically increasing spatial extent of the anomalous regime,
rather than as a divergence of the dimensionless crossover time.
At relatively short time scales,
$r_{\rm typ}(t)\sim(\Gamma_0t)^{(\mu-1)/2}$; thus,
$\mu-1$ provides the limiting small value of the effective
mean-square-displacement exponent $\xi(t)$, although reaching this
limit becomes increasingly difficult for small $\mu-1$.

\begin{figure}
     \centering
     \includegraphics[width=\linewidth]{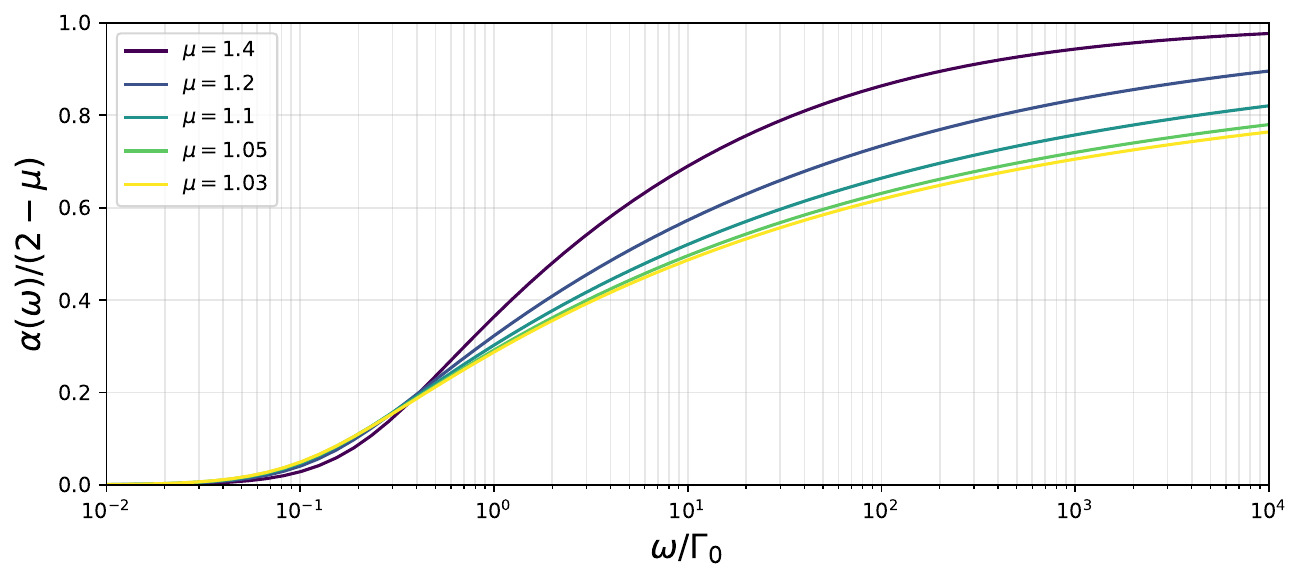}
     \caption{Frequency-dependent conductivity scaling exponent $ \alpha(\omega) \equiv \frac{d\ln\sigma}{d\ln\omega}$  for several values of the heavy-tail exponent $\mu$. The extended intermediate-frequency regime exhibits an apparent power-law behavior, while the low-frequency limit approaches the diffusive dc conductivity.}
     \label{Fig3}
 \end{figure}

 \textit{Frequency-dependent conductivity.--} Subdiffusive dynamics at moderately long times is naturally related,
 in frequency space, to the real part of \textit{ac} conductivity growing with frequency at $\omega \geq \Gamma_0$.
 This dependence can be found using 
 \eqref{conductivity main}, and is similar to a power law  $\sigma(\omega) \sim \omega^{\tilde\alpha}$, with 
 $\tilde\alpha < 1$. However, $\tilde\alpha$ is itself a slowly varying function of $\omega$; therefore, we present
 the result in the differential form
$ \alpha(\omega) \equiv \frac{d\ln\sigma}{d\ln\omega}$
where the dependence $\alpha(\omega)$ is shown in Fig.~\ref{Fig3}. In the high-frequency side, it has the following asymptotic form:
 \begin{equation}
     \frac{\alpha(\omega)}{2-\mu}\approx 
    1 - \frac{2\Gamma(\mu)}{\Gamma\left( 2 - \frac{2}{\mu}\right)|\omega/\Gamma_0|^{\mu-1}}  
 \end{equation}
 As $\mu\to1$, the regime of visibility of the above asymptotics shrinks.

\begin{figure}
    \centering
    \includegraphics[width=\linewidth]{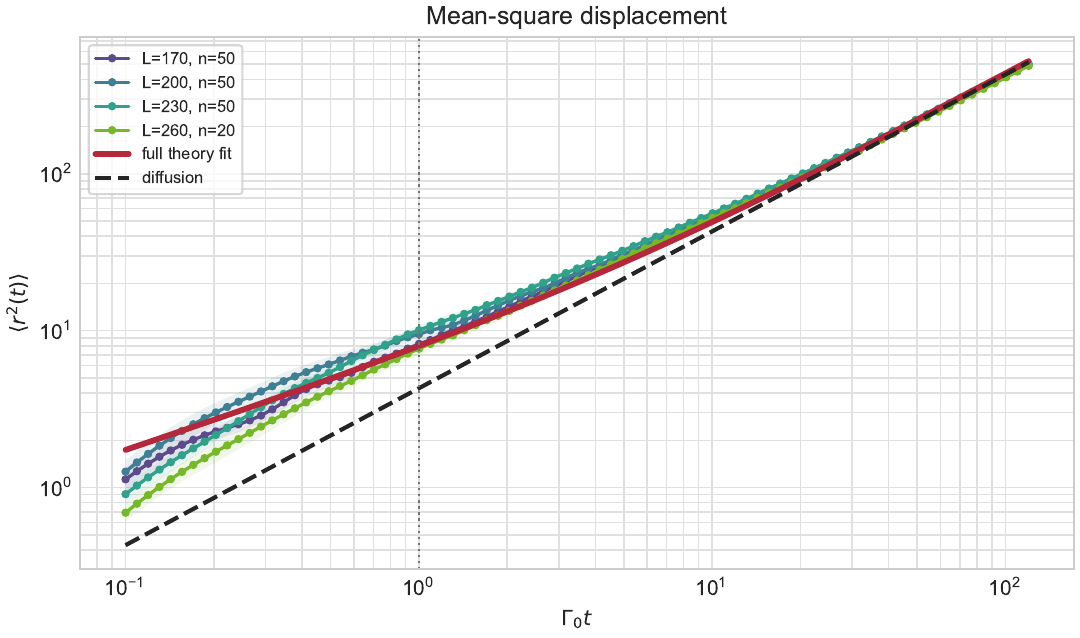}
    \caption{
    Mean-square displacement $\langle r^2(t)\rangle$ versus rescaled time
    $\Gamma_0 t$ for the unitary ($\beta=2$) power-law hopping model in 3D,
    Eq.~\eqref{anderson hamiltonian}, with $\mu=1.6$, $W=1$, $r_0=4.5$,
    $p_0=200$, and $h_0=0.07$, for several system sizes $L$, averaged over $n$ disorder realizations.
    The simulation data are compared with the full prediffusive theory obtained from Eqs.~\eqref{K func final}--\eqref{r^2} and with the asymptotic diffusive
    law $\langle r^2(t)\rangle=6D t$.
    The short- and intermediate-time dynamics exhibits a preasymptotic enhancement
    over diffusion and crosses over to diffusive scaling at long times. 
    }
    \label{Fig.4}
\end{figure}

\textit{Direct numerical simulations.--}We now compare our analytical predictions with direct numerical simulations of the Anderson model~\eqref{anderson hamiltonian}. We compute the mean-square displacement
\begin{equation}\label{numeric r2}
    \langle r^2(t)\rangle =
    \langle \psi(t)|
    \sum_n (\mathbf{r}_n-\overline{\mathbf{r}}_n)^2
    |\psi(t)\rangle ,
\end{equation}
where $|\psi(t)\rangle=e^{-i{\cal H}t}|\psi(0)\rangle$ and ${\cal H}$ is given by
Eq.~\eqref{anderson hamiltonian}. We consider a three-dimensional cubic lattice of
linear size $L$ with periodic boundary conditions in the unitary symmetry class
($\beta=2$). The initial state is localized at the center of the system, and the
diagonal disorder is independently drawn from a box distribution of width $W=1$.
Figure~\ref{Fig.4} shows the results for $\mu=1.6$.

A subtlety concerns the scale $\Gamma_0$ used to rescale time. At finite $h_0/W$, $\Gamma_0$ differs from its asymptotic result \eqref{Gamma}, derived in the semiclassical limit $h_0/W \ll 1$, [see Sec. 5.1 of
Ref. [29]]. We therefore determine $\Gamma_0$ directly from the long-time
diffusive asymptotics by fitting the numerical data to the linear
form $\langle r^2(t)\rangle = 6D t$, using \eqref{D coef} and extracting the corresponding
rescaling factor.

We note that direct simulations close to $\mu=1$ become increasingly
demanding, since the crossover length grows rapidly while maintaining
the controlled analytical regime requires progressively more restrictive
model parameters.

\textit{Conclusions.--} We have shown that broadly distributed hopping amplitudes can generate a pronounced regime of transport with a subdiffusive effective exponent in a noninteracting disordered system, while ordinary diffusion is ultimately restored at long times. As the localization transition is approached, the spatial extent of this anomalous regime grows parametrically, allowing subdiffusive-like dynamics to persist over distances much larger than the microscopic hopping scale. The effect originates entirely from disorder and rare hopping processes and therefore does not require interactions or genuine many-body mechanisms. The analytical predictions are consistent with direct numerical simulations at numerically accessible parameters. Although the resulting phenomenology is reminiscent of Griffiths-type dynamics near the many-body localization transition, the connection is only qualitative and does not imply the existence of a Griffiths phase. Our results provide a simple analytical benchmark for identifying disorder-induced transport anomalies and for separating them from effects that rely essentially on many-body physics.

\textit{Acknowledgments.} We are grateful to Vladimir Kravtsov and Yan Fyodorov for numerous useful discussions during this project. 

\bibliographystyle{unsrt}
\bibliography{bib}

\begin{widetext}
\begin{center}
{\Large\bfseries End Matter}
\end{center}
\end{widetext}

\section{Outlining the calculation}\label{Sec: calculation}

\textit{Supersymmetric formalism.--} In this End Matter, we briefly summarize the supersymmetric field-theoretical formulation following Ref.~\cite{FyodorovMirlinSommers}, using a notation common to the orthogonal ($\beta=1$) and unitary ($\beta=2$) ensembles.

We introduce commuting (bosonic) variables $\phi_n$ and anticommuting (Grassmann) variables $\chi_n$ satisfying
\begin{equation}
(\chi^\dagger)^\dagger=-\chi,\qquad
\chi_i\chi_j=-\chi_j\chi_i,\qquad
\int d\chi\,\chi=\frac{1}{\sqrt{2\pi}},
\end{equation}
together with the Grassmann representation of the determinant of an $N\times N$ matrix $\hat{A}$
\begin{equation}
\det\hat A=
\int
\frac{d\boldsymbol{\chi}\,d\boldsymbol{\chi}^\dagger}
{(2\pi)^{N/2}}
\exp\!\left(
\boldsymbol{\chi}^\dagger
\hat A
\boldsymbol{\chi}
\right).
\end{equation}
Using the identity $\text{Tr}\ln \hat{A} = \ln\det\hat{A}$, the relation above, and the property that $ \left[\hat{A}^{-1}\right]_{jk}=\frac{\partial}{\partial A_{jk}}\text{Tr}\ln\hat{A}$, one can obtain the relation $(\hat A^{-1})_{jk}=
\int
\frac{d\boldsymbol{\chi}\,d\boldsymbol{\chi}^{\dagger}}
{(2\pi)^{N/2}}
\chi_j^\dagger\chi_k\,
e^{\boldsymbol{\chi}^{\dagger}\hat A\boldsymbol{\chi}}/\det \hat{A},$ which allows us to represent the correlation function in Eq.~\eqref{K func} in the following form:
\begin{equation}
K(j,k)=
\int[d\psi]\,
\chi_{2j}^{\dagger}\chi_{1j}\chi_{2k}\chi_{1k}^{\dagger}\,
\overline{e^{S[\psi]}},
\label{eq:K_super}
\end{equation}
where the action reads
\begin{equation}
S[\psi]=
\frac{i\beta}{2}\sum_{n,m}
\psi_n^\dagger
\hat L
\left[
\left(E+\frac{i\kappa}{2}\hat L\right)\delta_{nm}
-H_{nm}
\right]
\psi_m,
\label{eq:action}
\end{equation}
The bosonic and fermionic fields are combined into retarded--advanced supervectors $\psi_n=
(\psi_R, \psi_A)^{T}$,
\begin{equation}\label{surervertor def}
\psi_{n}=\left(\begin{array}{c}
\psi_{Rn}\\
\psi_{An}
\end{array}\right),\quad\begin{array}{c}
\psi_{R/A}^{(\beta=1)}=\left(\begin{array}{cccc}
\phi_{R/A}^{(1)} & \phi_{R/A}^{(2)} & \chi_{R/A} & \chi_{R/A}^\dagger\end{array}\right)^T\\
\psi_{R/A}^{(\beta=2)}=\left(\begin{array}{cc}
\phi_{R/A} & \chi_{R/A}\end{array}\right)^T
\end{array},
\end{equation}
where $\phi_{R/A}^{(1),(2)}$ are real numbers, representing the components of the vector $\vec{\phi}_{R/A}$, while $\phi_{R/A}$ is a complex variable. The integration measure is $[d\psi]=
\prod_{n=1}^{N}
[d\psi_{Rn}\,d\psi_{Rn}^\dagger]
[d\psi_{An}\,d\psi_{An}^\dagger]$
\[
[d\psi_{An}\,d\psi_{An}^\dagger]=\begin{cases}
\frac{d\vec{\phi}d\chi d\chi^{\dagger}}{2\pi} & \beta=1\\
\frac{d\phi d\chi}{\sqrt{2\pi}} & \beta=2
\end{cases}
\]
and $\hat L=\mathrm{diag}(\mathbb{1},-\mathbb{1})_{R/A}$ distinguishes the retarded and advanced sectors. The overline denotes averaging over the random Hamiltonian $\hat H$ defined after Eq.~\eqref{K func}, and the sum runs over a macroscopic number of sites.

\textit{Averaging over off-diagonal disorder.--}
We begin by averaging over the off-diagonal disorder. The characteristic scale of the on-site disorder, $W$, is assumed to be much larger than the typical Lévy hopping amplitude, $W/h_0 \gg 1$. Consequently, the averages over the on-site disorder distribution $P_D(u)$ and the hopping distribution ${\cal P}(t)$ decouple. The action can therefore be written as
\begin{equation}\label{S action}
\begin{aligned}
S[\psi]
=
\frac{i\beta}{2}\sum_n \psi_n^\dagger \left((E-u_n)\hat{L} + \frac{i\kappa}{2}\right)\psi_n
+ \\ \sum_{m<n} w(\mathbf{r}_n-\mathbf{r}_m){\cal I}\left(\psi_n^\dagger\hat{L}\psi_m\right),
\end{aligned}
\end{equation}
where $w(\mathbf{r}_n-\mathbf{r}_m) {\cal I}\left(\psi_n^\dagger\hat{L}\psi_m\right) \equiv
\ln\left\langle e^{-x_{nm}}\right\rangle_{{\cal P}(t)}$,
and $
x_{nm}
=
\frac{i\beta}{2}
\left(
\psi_n^\dagger\hat{L}\psi_m t_{nm}
+
\psi_m^\dagger\hat{L}\psi_n t_{mn}
\right).
$ The hopping distribution ${\cal P}(t)$ was defined in Eq.~\eqref{hopping distr}. The restriction $m<n$ reflects the fact that, in the symmetry classes considered here, $t_{nm}$ and $t_{mn}$ are not statistically independent. For $\beta=1$, one has
$\psi_n^\dagger\hat{L}\psi_m t_{nm}= \psi_m^\dagger\hat{L}\psi_n t_{mn}$, whereas for $\beta=2$,
$\psi_n^\dagger\hat{L}\psi_m t_{nm}=\psi_n^\dagger\hat{L}\psi_m t_{mn}^{*}$.

Using the distribution ${\cal P}(t)$ from Eq.~\eqref{hopping distr}, we obtain
\begin{equation}
{\cal I}\left(\zeta_{nm}\right)
=
\frac{\ln\left[
1+\tilde{p}\,
w(\mathbf{r}_n-\mathbf{r}_m)
\left\langle e^{-x_{nm}}-1\right\rangle_{P_\mu(t)}
\right]}{w(\mathbf{r}_n-\mathbf{r}_m)}.
\end{equation}
The spatial dependence has thus been separated, leaving only the average over the Lévy distribution $P_\mu(t)$, whose form is given in the table following Eq.~\eqref{hopping distr}. The leading contribution to the integral $\left\langle e^{-x_{nm}}-1\right\rangle_{P_\mu(t)}$ follows from
\begin{equation}\label{Levy average integral}
\int_0^\infty d|t|\,P_\mu(|t|)e^{-z|t|}
\simeq
1+
\frac{h_0^\mu\mu \Gamma(-\mu)}{p_0\Gamma\left(1-\mu/2\right)}
z^\mu,
\end{equation}
derived in Ref.~\cite{LunkinTikhonov}; see Supplemental Material, Sec.~A.1. This approximation holds because $p_0$ is large, which guarantees small corrections. Since $\tilde{p}/p_0 = w_0^{-1}$, the parameter $[h_0 z]^{\mu}/w_0$ controls the logarithmic expansion. Its smallness is demonstrated below at the end of the Section \textit{Saddle-point equation}: ($z\sim \psi^\dagger \hat{L}\psi \sim W/h_0^2$). Importantly, this condition does not require the probability $\tilde{p}$ itself to be small. For the two symmetry classes, the result is
\begin{equation}\label{cal I def}
{\cal I}\left(\zeta_{nm}\right)
=
\begin{cases}
\displaystyle
\frac{h_0^\mu
\Gamma\left(\frac{\mu}{2}+1\right)}
{w_0 \Gamma(\mu+1)}
\left|\psi_n^\dagger\hat{L}\psi_m \right|^\mu,
& \beta=1,
\\[8pt]
\displaystyle
\frac{h_0^\mu}
{w_0\Gamma\left(\frac{\mu}{2}+1\right)}
\left[\psi_n^\dagger\hat{L}\psi_m  \psi_m^\dagger\hat{L}\psi_n \right]^{\mu/2},
& \beta=2.
\end{cases}
\end{equation}
More details are provided in Sec.~B.1 of the Supplemental Material of Ref.~\cite{DensCorrLevy} for $\beta=2$ and in Eqs.~(10)--(15) of Ref.~\cite{SpectralLRP} for $\beta=1$.

\textit{Results of the supersymmetric calculation.--}
Here we summarize the results of supersymmetric calculation developed in detail in Ref.~\cite{FyodorovMirlinSommers}; see Eqs.~(1)--(54) (except (32)-(39)), with the correspondence
\(\Gamma(\phi_n^\dagger\hat L\phi_m)\leftrightarrow{\cal I}(\zeta_{nm})\).
Following this procedure, the problem reduces to equations involving only commuting variables, analogous to Eqs.~(48)--(54) of Ref.~\cite{FyodorovMirlinSommers}. In particular,
\begin{equation}\label{Kqsum}
\tilde K_q(\kappa)
=
\sum_\nu
\frac{c_\nu(\kappa)}
{1-\tilde w(q)\lambda_\nu(\kappa)},
\qquad
c_\nu(\kappa)
=
\frac{(f_\nu,1)^2}{(f_\nu,f_\nu)}.
\end{equation}
The scalar product is defined as
\begin{equation}\label{scalar prod}
(f_1,f_2)
\equiv
\int\frac{d\phi_R\,d\phi_A}{(2\pi)^2}
F(\phi, \kappa)e^{g_\kappa(\phi)}
f_1(\phi)f_2(\phi),
\end{equation}
where
\begin{equation}\label{F(phi) def}
F(\phi, \kappa)
=
\int du\,P_D(u)
\exp\left[
\frac{i\beta}{2}(E-u)\phi^\dagger\hat L\phi
-\frac{\beta\kappa}{2}\phi^\dagger\phi
\right].
\end{equation}
Here, vectors \(\phi\) and \(\phi^\dagger\) contain the commuting components of the supervector defined in Eq.~\eqref{surervertor def}:
\begin{equation}
\begin{aligned}
\phi^{(\beta=1)}
&=
\left(
\phi_R^{(1)},
\phi_R^{(2)},
\phi_A^{(1)},
\phi_A^{(2)}
\right)^T,
\\
\phi^{(\beta=2)}
&=
\left(
\phi_R,
\phi_A
\right)^T.
\end{aligned}
\end{equation}
The function \(g_\kappa(\phi)\) is the projection of the saddle-point solution to the commuting sector and satisfies
\begin{equation}\label{saddle point eq}
g_\kappa(\psi)
=
w_0\int[d\psi']
\,{\cal I}(\psi^\dagger\hat L\psi')
F(\psi')e^{- g_\kappa(\psi')}.
\end{equation}
For \(\beta=2\), this equation was solved in detail in Sec.~C.2 of the Supplemental Material of Ref.~\cite{DensCorrLevy}. For \(\beta=1\), the calculation and approximations remain unchanged, apart from the modified kernel in Eq.~\eqref{cal I def}; the corresponding derivation is presented in the following Section. To leading order in \(h_0/W\), the real part of solution of Eq.~\eqref{saddle point eq} is
\begin{equation}\label{g(phi)}
g_\kappa(\phi)
\simeq
\displaystyle
\left[
\frac{\beta \Gamma_\kappa}{4}\phi^\dagger\phi
\right]^{\mu/2},
\end{equation}
where \(\Gamma_\kappa\) is defined in Eq.~\eqref{Gamma}.

In Eq.~\eqref{Kqsum}, \(f_\nu\) is the eigenfunction associated with the eigenvalue \(\lambda_\nu\) of the following integral equation, obtained after integrating out the Grassmann variables:
\begin{equation}\label{integral eq 1}
\int\frac{d\phi}{(2\pi)^2}
F(\phi, \kappa)e^{g_\kappa(\phi)}
J(\phi,\phi')f_\nu(\phi)
=
\lambda_\nu (\kappa) f_\nu(\phi').
\end{equation}
The kernel \(J(\phi,\phi')\) depends on the symmetry class:
\begin{equation}\label{different J kernels}
\begin{aligned}
J^{(\beta=1)}(\phi,\phi')
&=
{\cal I}''(\phi^\dagger\hat L\phi'),
\\
J^{(\beta=2)}(\phi,\phi')
&=
{\cal I}''(\phi^\dagger\hat L\phi')
(\phi'^\dagger\hat L\phi \phi^\dagger\hat L\phi')
+
{\cal I}'(\phi^\dagger\hat L\phi').
\end{aligned}
\end{equation}
Equation~\eqref{Kqsum} coincides with Eq.~\eqref{K func final} because of the specific form of the functions $f_{\nu}$, as explained in detail below in the Section \textit{Derivation of main results}.

\textit{Saddle-point equation.--} In this section, we derive \eqref{g(phi)}
from Eq.~\eqref{saddle point eq}. We seek a solution depending on two invariants, of the form $ g_\kappa(\psi) = \tilde{g}_\kappa(|\psi_R|^2,|\psi_A|^2)$. We first integrate out the Grassmann variables by expanding  Eq.~\eqref{saddle point eq} using the formula

\begin{widetext}
\begin{equation}
 f(|\psi_R|^2,|\psi_A|^2) =  f(|\phi_R|^2,|\phi_A|^2) +\frac{2}{\beta}\chi_R^*\chi_R\,\frac{\partial f(|\phi_R|^2,|\phi_A|^2)}{\partial |\phi_R|^2}  
 +\frac{2}{\beta}\chi_A^*\chi_A\,\frac{\partial f(|\phi_R|^2,|\phi_A|^2)}{\partial |\phi_A|^2}  +\frac{4}{\beta^2}\chi_R^*\chi_R\chi_A^*\chi_A\,\frac{\partial^2 f(|\phi_R|^2,|\phi_A|^2)}{\partial |\phi_R|^2\partial |\phi_A|^2} .
\label{eq:grassmannians_expansion}
\end{equation}
\end{widetext}
so that, after integration over the Grassmann variables, \eqref{saddle point eq} becomes
\begin{equation}\label{saddle point eq real part}
    g_\kappa(\phi)
=
\frac{4w_0}{\beta^2}\int\frac{d\phi'}{(2\pi)^2}
\,{\cal I}(\phi^\dagger\hat L\phi')\frac{\partial^2 \left[ F(\phi', \kappa)e^{-g_\kappa(\phi')}\right]}{\partial |\phi_R'|^2\partial|\phi_A'|^2}.
\end{equation}
$F(\phi, \kappa)$ contains diagonal disorder $P_D(u)$. Since it is assumed to be a slowly varying function, we estimate 
\begin{equation}
\int du P_D(u)
 e^{\frac{i\beta}{2}(E-\zeta)(|\phi_R|^2-|\phi_A|^2)}
 \simeq
 \frac{4\pi P_D(E)}{\beta}\delta(|\phi_R|^2-|\phi_A|^2).
\label{delta-f}
\end{equation}
After integration by parts in Eq.~\eqref{saddle point eq real part}, the boundary terms vanish, yielding
\begin{equation}\label{saddle point eq partial int}
    \tilde{g}_\kappa(s,t)
=
\frac{4w_0 \pi P_D(E)}{\beta}\int ds' {\cal C}_\mu (s',s,s',t)
e^{-\frac{\beta \kappa s'}{2}-\tilde{g}_\kappa(s',s')}.
\end{equation}
with 
\begin{equation}
    {\cal C}_\mu(s',s,t',t) = \int \frac{d\gamma_R'd\gamma_A'}{(2\pi)^2} \left(\frac{\partial^2 {\cal I}(\phi^\dagger\hat L\phi')}{\partial |\phi_R'|^2\partial|\phi_A'|^2}\right)\biggr|_{\begin{array}{c} |\phi_R'|^2 = s' \\
     |\phi_A'|^2 = t'  \end{array}}
\end{equation}
Here we use the following set of variables:
\begin{equation}\label{s t gamma}
\int\frac{\left[d\phi\right]}{\left(2\pi\right)^{2}}=\frac{\beta^2}{4}\int\frac{ dsdtd\gamma_R d\gamma_A}{(2\pi)^2},\halfquad \begin{array}{c} t=\left|\phi_{A}\right|^{2}, \\ s=\left|\phi_{R}\right|^{2},\end{array}
\begin{array}{c}
     \frac{\phi_R}{\phi_R'}=
 \sqrt{\frac{s}{s'}}e^{i\gamma_R}\\
     \frac{\phi_A}{\phi_A'}=
 \sqrt{\frac{t}{t'}}e^{i\gamma_A} 
\end{array}.
\end{equation}
This set is universal for both symmetries, with the following representations:
\[
\begin{array}{c}
\text{Re}(\phi_{R/A})\big|_{\beta=2}= \phi^{(1)}_{R/A}\big|_{\beta=1}= |\phi_{R/A}|\cos(\gamma_{R/A}) \\
\text{Im}(\phi_{R/A})\big|_{\beta=2}= \phi^{(2)}_{R/A}\big|_{\beta=1}= |\phi_{R/A}|\sin(\gamma_{R/A})
\end{array}
\]
Within approximation~\eqref{delta-f}, only $g_\kappa(s,s)$ is needed for the subsequent calculations. It is therefore sufficient to solve the self-consistent equation~\eqref{saddle point eq partial int} for $t=s$. In this case, ${\cal C}_\mu$ from Eq.~\eqref{saddle point eq partial int} takes the form
\begin{equation}
\mathcal{C}_{\mu}(s',s,s',s)
=
\frac{
2^{\,\mu(\beta-1)-2}\,
\,h_0^\mu\,
s^{\mu/2}(s')^{\mu/2-2}\,
\Gamma\!\left(\frac{\mu-1}{2}\right)
}{w_0
\sqrt{\pi}\,
\Gamma\!\left(\frac{\mu}{2}\right)
\Gamma\!\left(\frac{\mu}{2}-1\right)
}.
\end{equation}
Substituting this expression into Eq.~\eqref{saddle point eq partial int} and using the definition of the Lévy stable distribution in Eq.~\eqref{Levy def}, we obtain the results present in Eq.~\eqref{Gamma}. 

The approximation~\eqref{delta-f} requires $|\phi^\dagger\hat L\phi|\gg  W^{-1}$.  This can be estimated as follows. Mathematically, the statistical independence of the real and imaginary parts
of the self-energy $\Sigma (E)$, discussed in the Section \textit{Physical discussion and regime of validity} of the Main text, follows from the approximate representation of $g(\phi)$ function. As shown in \cite{LunkinTikhonov}, it separates into two parts
\begin{equation}
    g(\phi)\approx g_{\text{Re}\Sigma}(\phi^\dagger \hat{L}\phi) + g_{\text{Im}\Sigma}(\phi^\dagger \phi),
\end{equation}
each of which controls the characteristic function (and thus the inverse typical scale) of the real and imaginary parts of the self-energy $\Sigma$, respectively. The fact that $\Re\Sigma(E)$ receives contributions from a wide energy range
of order $W$, while $\Im\Sigma(E)$ is governed only by the local environment of the level at energy $E$ follows from different typical values of $g_{\text{Re}\Sigma}(\phi^\dagger \hat{L}\phi)$ and $g_{\text{Im}\Sigma}(\phi^\dagger \phi)$. We can estimate $\phi^\dagger \hat{L}
\phi$ from Ref.~\cite{SpectralLRP}, where $g_{\text{Re}\Sigma}(\phi^\dagger \hat{L}\phi)$ was estimated. 
\begin{equation}
g_{\text{Re}\Sigma}(\phi^\dagger\phi) \sim 
 \frac{h_0^\mu }
 {W^{\mu/2}}|\phi^\dagger \hat{L} \phi |^{\mu/2}, \quad \phi^\dagger \hat{L}\phi \sim \frac{W}{h_0^2} \gg \frac{1}{W},
\end{equation}
which justifies Eq.~\eqref{delta-f} in the regime of interest. 

\textit{Derivation of main results.--} In this Section, we prove Eq.~\eqref{K func final} and derive Eqs.~\eqref{K func final}--\eqref{D coef} using the results summarized in the Section~\textit{Results of the supersymmetric calculation}. We focus on the integral equation~\eqref{integral eq 1}.  To evaluate it, we use the expressions introduced above: \eqref{g(phi)} for $g_\kappa(\phi)$,  Eqs.~\eqref{different J kernels} and \eqref{cal I def} for the kernel $J(\phi,\phi')$,  \eqref{F(phi) def}
with approximation \eqref{delta-f} for $F(\phi, \kappa)$ and universal coordinates \eqref{s t gamma}.

Combining these expressions, we find that the phase dependence enters only through $\cos(\gamma_R-\gamma^\prime_R)$ and $\cos(\gamma_A-\gamma^\prime_A)$. The eigenfunctions therefore have the form
\begin{equation}
{f}^{(\beta)}_{\nu}\left(s,t,\theta\right)={\cal F}^{(\beta)}_{\nu}\left(s,t\right)e^{i\nu_R\gamma_R + i\nu_A \gamma_A}
\end{equation}
The approximation~\eqref{delta-f} multiplies the kernel $J(\phi,\phi')$, so that \eqref{integral eq 1} takes the form 
\begin{equation}
    k^{(\beta)} \int_0^\infty  ds  s^{\frac{\mu}{2}-1}{\cal F}_\nu^{(\beta)} (s,s) e^{-\frac{\beta s \kappa}{2}-\left[\frac{\beta \Gamma_\kappa}{2} s\right]^{\mu/2}}   = \lambda_\nu (\kappa),
\end{equation}
where \begin{equation}
{\cal F}^{(\beta)}_\nu (s,t) = \int \frac{d\gamma_R d\gamma_A}{(2\pi)^2} e^{i\nu_R \gamma_R + i\nu_A \gamma_A} {\cal Y}^{(\beta)}(s,t, \gamma_R, \gamma_A),
\end{equation}
 with
\begin{equation}
\begin{aligned}
{\cal Y}^{(\beta = 1)}(s,t, \gamma_R, \gamma_A)  = 
    \left(\sqrt{s}\cos(\gamma_R) - \sqrt{t}\cos(\gamma_A)\right)^{\mu-2} \\
   {\cal Y}^{(\beta = 2)}(s,t, \gamma_R, \gamma_A)  =   \left( s + t -2\sqrt{st}\cos(\gamma_R - \gamma_A)\right)^{\frac{\mu}{2}-1} 
\end{aligned}
\end{equation}
and \begin{equation}
        k^{(\beta=1)} = \frac{\pi P_D (E) h_0^\mu \Gamma\left( \frac{\mu}{2}\right)\mu}{2 w_0 \Gamma\left(\mu-1\right) },\halfquad    k^{(\beta=2)} = \frac{\pi P_D (E) h_0^\mu\mu}{w_0 \Gamma\left(\frac{\mu}{2}\right) } .
\end{equation}

Returning to Eqs.~\eqref{scalar prod} and \eqref{Kqsum}, we see that because of the $\sim e^{i\nu_R \gamma_R + i\nu_A \gamma_A}$ dependence, the combination $c_\nu(\kappa) = \frac{(1, f_\nu)^2}{(f_\nu, f_\nu)}$  becomes zero for all $\nu$ except $\nu=0$. This proves Eq.~\eqref{K func final}; henceforth, we consider only the $\nu=0$ case. We therefore obtain the expressions for ${\cal F}^{(\beta)}_0(s,s)$:
\begin{equation}
    {\cal F}^{(\beta)}_0(s,s) = \begin{cases}
        s^{\frac{\mu}{2}-1} \frac{2^{\mu - 2}\left[\Gamma\left( \frac{\mu-1}{2}\right)\right]^2}{\pi \left[\Gamma\left(\frac{\mu}{2}\right)\right]^2}  & \beta = 1\\
        s^{\frac{\mu}{2}-1} \frac{2^{\mu-2}\Gamma\left( \frac{\mu-1}{2}\right)}{\sqrt{\pi} \Gamma\left(\frac{\mu}{2}\right)} & \beta =2 
    \end{cases}
\end{equation}
Using the definition in Eq.~\eqref{Levy def}, we obtain $\lambda_{0}(\kappa)$ and $c_0(\kappa)$:
\begin{eqnarray}
\lambda_{0}(\kappa)= k^{(\beta)} \int_0^\infty  ds \int dxL_{\mu/2}(x) s^{\frac{\mu}{2}-1} \times \\ \nonumber
{\cal F}_0^{(\beta)} (s,s) e^{-\frac{\beta s \kappa}{2}-\frac{\beta \Gamma_\kappa}{2} sx}  ,
\end{eqnarray}
\begin{equation}
\frac{c_0(\kappa)}{\beta\pi P_D(E)} = \frac{\left[\int ds\int dxL_{\mu/2}(x) e^{-\frac{\beta s \kappa}{2}-\frac{\beta \Gamma_\kappa}{2} sx} {\cal F}_0^{(\beta)}(s,s) \right]^2}{\int ds\int dxL_{\mu/2}(x) e^{-\frac{\beta s \kappa}{2}-\frac{\beta \Gamma_\kappa}{2} sx} \left[{\cal F}_0^{(\beta)}(s,s)\right]^2}
\end{equation}
After straightforward algebra and use of Eq.~\eqref{Gamma}, we obtain the final expressions in Eq.~\eqref{lambda c defs}. These results are common to both symmetry classes $\beta=1,2$. They have the following asymptotics:
\begin{itemize}
    \item $\kappa \ll \Gamma_0$
\begin{equation}
    \lambda_{0}(\kappa )\approx\frac{1 - \frac{w_2\kappa}{w_0 D}}{w_{0}},\halfquad \frac{c_0(\kappa )}{2\pi P_D (E)} \approx \frac{ w_2\left(1 + \frac{\mu^2-2}{(\mu-1)(\mu-2)} - \frac{4\pi/\mu}{\sin\left(\frac{2\pi}{\mu}\right)}\right)}{w_0 D},
\end{equation}
    \item $\kappa \gg\Gamma_0$
\end{itemize}
\begin{equation}
\lambda_0(\kappa)  \approx\frac{\mu \Gamma\left(\mu-1\right)\left[\frac{\Gamma_{0}}{\kappa}\right]^{\mu-1}}{2w_0\Gamma\left(2-\frac{2}{\mu}\right)},\halfquad c_0(\kappa) \approx \frac{2\pi P_D (E) \left[\Gamma\left(\frac{\mu}{2}\right)\right]^2}{\kappa \Gamma(\mu-1)}.
\end{equation}
Equation~\eqref{correction <r2>} can be derived from a higher-order small-$\kappa$ expansion of Eq.~\eqref{lambda c defs}. Substitution into Eq.~\eqref{r^2} then yields the  result shown in Eq.(\eqref{correction <r2>}).

\end{document}